\documentclass[useAMS,usenatbib]{aastex}

\usepackage{times}
\usepackage{graphicx} 
\usepackage[export]{adjustbox}
\usepackage{spr-astr-addons}
\usepackage{url}
\usepackage{hyperref}
\usepackage{rotating}
\usepackage[caption=false]{subfig}
\usepackage{aas_macros}
\usepackage{tabularx}
\usepackage{multicol}
\usepackage[usenames]{color} 
\usepackage{sidecap}
\usepackage{textcomp}

\renewcommand{\thefootnote}{\fnsymbol{footnote}}
\newcommand{\msun}{\mathrm{M}_\odot} 
\newcommand{\rsun}{\mathrm{R}_\odot} 
 
\newcommand{\kms}{km~s\ensuremath{^{-1}}}
\newcommand{\cycperday}{d\ensuremath{^{-1}}}
\newcommand{\cpd}{d\ensuremath{^{-2}}}
\newcommand{\teff}{$T_{\rm eff}$} 
\newcommand{\logg}{$\log g$} 
\newcommand{\vsini}{$v \sin i$} 
\newcommand{\logL}{$\log (L/L_\odot)$}
\newcommand{\logT}{$\log T_{\rm eff}$}

\newcommand{\gmodes}{$g$-modes}
\newcommand{\pmodes}{$p$-modes}
\newcommand{\logamp}{$\log (Amp)$}
\newcommand{\logen}{$\log (En)$}
\newcommand{\logeff}{$\log (Eff)$}
\newcommand{\bvo}{$(B-V)_0$}
\newcommand{\emaila}{sowgata.edu@gmail.com}
\begin{document}

\title{\bf{A study of Pulsation \& Rotation in a sample of A-K type stars in the {\it Kepler} field}}
\shorttitle{Pulsation and Rotation in A-K type stars in the {\it Kepler} field}
\shortauthors{Sowgata et al.}

\author{Sowgata Chowdhury\altaffilmark{1,2,3}} 
\email{\emaila}

\author{Santosh Joshi\altaffilmark{3}}

\author{Chris A. Engelbrecht\altaffilmark{4}}

\author{Peter De Cat\altaffilmark{5}}

\author{Yogesh. C. Joshi\altaffilmark{3}}

\author{K. T. Paul\altaffilmark{2}}

\altaffiltext{1}{Nicolaus Copernicus Astronomical Center, Bartycka 18, PL-00-716 Warsaw, Poland}

\altaffiltext{2}{Department of Physics, CHRIST (Deemed to be University), Hosur Road, Bengaluru - 560029, Karnataka, India}

\altaffiltext{3}{Aryabhatta Research Institute of Observational Sciences (ARIES) Manora peak, Nainital - 263002, India}

\altaffiltext{4}{Department of Physics, University of Johannesburg, PO Box 524, Auckland Park, Johannesburg 2006, South Africa}

\altaffiltext{5}{Royal Observatory of Belgium Ringlaan 3, B-1180 Brussel, Belgium}

\def\LaTeX{L\kern-.36em\raise.3ex\hbox{a}\kern-.15em 
 T\kern-.1667em\lower.7ex\hbox{E}\kern-.125emX} 
 
\begin{abstract}

We present the results of time-series photometric analysis of 15\,106 A-K type stars observed by the {\it Kepler} space mission. 
We identified 513 new rotational variables and measured their starspot rotation periods as a function of spectral type and discuss the distribution of their amplitudes. We examined the well-established period-colour relationship that applies to stars of spectral types F5-K for all of these rotational variables and, interestingly, found that a similar period-colour relationship appears to extend to stars of spectral types A7 to early-F too. This result is not consistent with the very foundation of the period-colour relationship. We have characterized 350 new non-radial pulsating variables such as A- and F-type candidate $\delta$\,Scuti, $\gamma$\,Doradus and hybrid stars, which increases the known candidate non-radial pulsators in the {\it Kepler} field significantly, by $\sim$20\%. The relationship between two recently constructed observables, $Energy$ and $Efficiency$, was also studied for the large sample of non-radial pulsators, which shows that the distribution in the logarithm of $Energy$ (\logen) can be used as a potential tool to distinguish between the non-radial pulsators, to some extent. Through visual inspection of the light curves and their corresponding frequency spectra, we found 23 new candidate red giant solar-like oscillators not previously reported in the literature. The basic physical parameters such as masses, radii and luminosities of these solar-like oscillators were also derived using asteroseismic relations.

\end{abstract}


\section{Introduction} \label{sec:intro}

In the past, many photometric and spectroscopic surveys were conducted in both the Northern and Southern hemisphere to search for low amplitude variability in A- to F-type stars  
\citep{Martinez1991,Nelson1993,Paunzen2012,Kochukhov2006,Kurtz2006}. 
At the Aryabhatta Research Institute of Observational Sciences (ARIES), Nainital, India, we initiated a similar research project to detect low-amplitude photometric variability and study it in more detail using spectroscopy \citep{Ashoka2000,Martinez2001,Joshi2003,Joshi2006,Joshi2009,Joshi2010,Joshi2012,Joshi2016,Joshi2017}.
 
The best precision ever obtained from ground-based photometric observations is 14 $\mu$mag \citep{Kurtz2005} which is not even sufficient to detect the amplitude variation generally found in solar-like stars. 
After the launch of various space missions, much lower detection levels of the order of $\mu$mag have been achieved: WIRE \citep{Buzasi2000}, MOST \citep{Matthews2004}, CoRoT \citep{Baglin2009}, and {\it Kepler} \citep{Chaplin2010,Koch2010}. 
This motivated us to extend our survey program from ground-based to space-based observations. This resulted in the formation of a new collaboration with South African astronomers to detect and study photometric variability using the {\it Kepler} archival data. 
The first two papers of this collaboration are published by \citet{Balona2013b} and \citet{Balona2016}. In this third paper, we study the time-series photometric data of 15\,106 A-K type stars observed by the {\it Kepler} space mission.

Space-based observations of variable stars have revolutionized the field of variability studies. 
Dedicated missions such as CoRoT and {\it Kepler} have duty cycles which are unachievable from the ground.
It effectively solves many problems prevalent in ground-based observation campaigns. 
In particular, the {\it Kepler} mission in its original operational mode collected data of unprecedented photometric precision for almost 200,000 stars in a large field of 105 square degrees in the direction of Cygnus and Lyra. 
In addition to its primary objective of detecting planetary transits, {\it Kepler} has provided high quality data for many areas of investigation, particularly asteroseismology and rotation-related activity. We have taken advantage of the photometric quality of data from the {\it Kepler} mission to study variable A-K type stars in the original {\it Kepler} field.

The long cadence mode (LC) of {\it Kepler} (one observation every 29.4 min) is perfectly adequate for detecting variability in red giants. Solar-like oscillations have been discovered in thousands of them \citep{Bedding2010, Huber2010,Kallinger2010a}, including some in open clusters \citep{Hekker2011,Stello2010, Basu2011}. 
It is possible to independently estimate the mass $M$, radius $R$ and surface gravity \logg\ of each star, in the case of pulsating red giants. In these stars, the mean density, mass and radius can be obtained from the large separation $\Delta\nu$ (i.e. frequency spacing between modes of consecutive overtones $n$ but of same spherical harmonic degree $\ell$) and frequency of maximum amplitude $\nu_{\rm max}$. \citet{Stello2010} was the first to apply this method to red giants in the cluster NGC\,6819, using {\it Kepler} data. 

In addition to its primary objective of detecting planetary transits, {\it Kepler}'s high quality data provide a valuable opportunity for specifically investigating relationships between several areas of research in stellar structure and evolution. 
The main sequence $\delta$\,Scuti ($\delta$\,Sct) and $\gamma$\,Doradus ($\gamma$\,Dor) stars with masses between 1.2 and 2.5 $\msun$ are particularly useful for pulsation studies, to infer the interior structure of stars and to test theoretical models describing them. 
The $\delta$\,Sct stars pulsate in pressure modes (\pmodes), giving us the opportunity to study the stellar envelope while the $\gamma$\,Dor stars pulsate in gravity modes (\gmodes), that help to study the regions near the stellar core. 
Hybrid objects are pulsating in both types of modes and are hence of great asteroseismic interest because they can give constraints on the full stellar structure. 
\citet{Debosscher2011} used automated algorithms to classify the variable stars in the {\it Kepler} field. 
On combining the first two quarters of data for hundreds of variable stars, \citet{2010ApJ...713L.192G} showed that the frequency spectra are so rich that there are practically no pure $\delta$\,Sct or $\gamma$\,Dor pulsators. 
Essentially all the stars in the corresponding area of the Hertzsprung-Russell (H-R) diagram show pulsational frequencies in both the $\delta$\,Sct and $\gamma$\,Dor frequency ranges. 
To categorize these non-radial pulsators, the authors proposed a new observational classification scheme, taking into account the amplitude of the dominant mode as well as the range of observed frequencies. 

This paper reports on the outcome of a study for which the key goals are threefold:

(i) to identify additional solar-like oscillators. For the solar-like oscillators found in our sample, we derive their basic parameters (e.g. mass, radius and luminosity) using the asteroseismic scaling relations. 

(ii) to explore the nature of a rotation period-colour relation (if any) that might extend blueward of the canonical limit at $(B-V)$ =\,0.47 (see \citealt{Mamajek2008}). 
This is motivated by \citet{Balona2013b,Balona2013a}, who found a continuation of the canonical period-colour relation to stars with $(B-V)$ $>$\,0.2, which demands a re-evaluation of the physical cause of this relationship (or a re-evaluation of the presence of convection in A-type stars).  We therefore searched for spotted stars in our sample to investigate this relationship. 
 
(iii) to search for other non-radial pulsators in our sample, using some automated constraints and independently confirming them by visual inspection of all the light curves and corresponding Discrete Fourier transforms (DFTs). Finding new non-radial pulsators is imperative to advance our understanding of such variable stars, through asteroseismology. 
Two recently constructed observables, $Energy$ and $Efficiency$ \citep{Uytterhoeven2011} have also been studied for our large sample of non-radial pulsators to explore the relationship between $\delta$\,Sct, $\gamma$\,Dor and candidate hybrid variables in the parameter space of these new observables. 

Apart from the DFT technique, the analysis of variance (AoV) method \citep{Czerny1989, Schwarzenberg1996} was also adopted to estimate the period of variability and verify the classification in all cases. We study the {\it Kepler} field stars using the Sloan {\it g,r,i,z} photometry retrieved from the {\it Kepler} Input Catalogue (KIC; \citealt{Brown2011}) and time-series photometry from the {\it Kepler} archive at the Mikulski Archive for Space Telescopes (MAST).

The work presented here is organized as follows: In Sec. 2 we give the sample selection and description of the {\it Kepler} data archive. In Sec. 3, 4 and 5 the different classes of pulsating and rotating stars are discussed, followed by the conclusion drawn from our study in Sec. 6. \\

\section{Data description}

The data presented in this work include all available LC data from quarter 0 (Q0) to quarter 17 (Q17) for 15\,106 {\it Kepler} light curves. 
We selected stars of spectral types ranging from A0 to K7 as this range included the cooler red giant stars where solar-like oscillations are likely to be present. 
The hotter stars in the spectral region $(B-V)$ $<$\,0.47 were also included for the investigation of the period-colour relationship in rotational variables. 
The complete list of stars that were studied in this paper is provided online.

For the vast majority of stars, {\it Kepler} photometry is available only in the LC mode, which consists of practically uninterrupted exposures with a cadence of 29.4 minutes. 
For a few thousand stars, the short cadence (SC) mode of $\sim$1-minute exposures is available, but these usually cover only one or two months of observations. 
{\it Kepler} observed in white light, with a passband with a FWHM covering the 430-890\,nm wavelength range. 
The observed stars have {\it Kepler} magnitudes ranging from 9 to 16 mag. 
The {\it Kepler} data are available in either ‘uncalibrated’ or ‘calibrated’ form. 
The calibrated data suffer from artefacts caused by the processing and are not used here. 
Details of the technique used to correct the uncalibrated data and how the corrections affect the derived frequencies can be found in \citet{Balona2011a}. As stated in \citet{Balona2011a}, the trend-removing procedure for Kepler data ``... tends to dampen or remove very low frequencies...'', such that, in practice, the lowest frequency that can be detected in these data is about 0.1 ~\cycperday\ . \citet{Balona2011a} go on to say ``but frequencies above about 0.1 ~\cycperday\ are not affected''. We took this as a vote of confidence in the determination of rotational periods shorter than 10 days and therefore confined our study to rotational periods below this upper limit. The Nyquist frequency of the Kepler LC data is around 24 ~\cycperday\ . All the data used in this paper are publicly available and can be retrieved using MAST.

\section{Solar-Like Oscillations} \label{sec:sol-like}

Solar-like oscillations can be expected in low-mass main-sequence stars, sub-giants, stars on the red-giant branch (RGB), horizontal branch and asymptotic-giant branch stars \citep{Christensen1983,Houdek1999,Dziembowski2001}. Following the success of helioseismology, the large volumes of long time-series data of exquisite quality, in particular from the CoRoT and {\it Kepler} space missions, have been particularly important for the study of solar-like oscillations in stars, owing to the small amplitudes of these oscillations. In particular, after the launch of {\it Kepler} in 2009, there has been tremendous progress in asteroseismology of G- and K-type red giants, with the detection of solar-like oscillations in thousands of stars. These oscillations are easily identified in the periodogram because of the localized comb-like structure where amplitudes decrease sharply from a central maximum. 

Using {\it Kepler} data, \citet{Huber2014} presented a revised catalog of 196,468 stars observed in Q1 to Q16 and found oscillations in 2762 giants, thereby increasing the number of known oscillating giant stars observed by $\sim$20\% (to $\sim$15,500 stars). 
With {\it Kepler}\textquotesingle s ecliptic second-life mission, 
K2 \citep{Howell2014}, many new pulsating stars with solar-like oscillations have been discovered, 
i.e. 4 sub-giants \citep{Chaplin2015}, 55 red-giants \citep{Stello2015}, 33 solar-type stars \citep{Lund2016a}, 2 main sequence stars of the Hyades open cluster \citep{Lund2016b}, 8 stars in the globular cluster M4 \citep{Miglio2016}, 33 red giants in M67 \citep{Stello2016} and 1210 further red giants \citep{Stello2017}.

\begin{table*}
\caption{A list of 23 new stars with solar-like oscillations discovered by visual inspection. Kp is the {\it Kepler} Magnitude. The first five columns are KIC parameters, except the second column, which is revised effective temperature from \citet{Pinsonneault2012}. The calculated parameters i.e., frequency of maximum amplitude, the large separation, mass, radius and luminosity calculated from the solar-like oscillations are shown from sixth to tenth column, respectively. The last two columns are stellar parameters from GAIA DR2. The complete table is available online.} 
\label{Sol_Like}
\begin{tabular}{cccccccccccc}
\hline
\multicolumn{1}{c}{KIC} & 
\multicolumn{1}{c}{\teff} & 
\multicolumn{1}{c}{$\log (L/$L$_{\odot})$} & 
\multicolumn{1}{c}{\logg} & 
\multicolumn{1}{c}{Kp} & 
\multicolumn{1}{c}{$\nu_{\rm max}$} &
\multicolumn{1}{c}{$\Delta\nu$} &  
\multicolumn{1}{c}{Mass} & 
\multicolumn{1}{c}{Radius} &
\multicolumn{1}{c}{$\log (L/$L$_{\odot})$} &
\multicolumn{1}{c}{Radius$_g$} &
\multicolumn{1}{c}{$\log (L/L_{\odot})_g $} 
\\
\multicolumn{1}{c}{(Id)} & 
\multicolumn{1}{c}{(K)} & 
\multicolumn{1}{c}{ } & 
\multicolumn{1}{c}{$(cm s^{-2})$} & 
\multicolumn{1}{c}{(mag)} & 
\multicolumn{1}{c}{$(\textmu $$Hz$$)$} &
\multicolumn{1}{c}{$(\textmu $$Hz$$)$} & 
\multicolumn{1}{c}{$(\msun)$} & 
\multicolumn{1}{c}{$(\rsun)$} &
\multicolumn{1}{c}{ } &
\multicolumn{1}{c}{$(\rsun)$} &
\multicolumn{1}{c}{ } \\ 
\hline
002311130 & 4822 & 1.909 & 2.306 & 11.932 & 18.143 & 2.357 & 1.607 & 17.398 & 2.170 &16.321 & 2.042 \\
002437851 & 4915 & 0.756 & 3.447 & 16.353 & 10.399 & 1.338 & 2.998 & 31.242 & 2.712 &   -   &   -   \\
002439630 & 4782 & 1.810 & 2.378 & 13.325 & 12.948 & 2.029 & 1.050 & 16.686 & 2.119 &17.927 & 2.055 \\
002568575 & 4811 & 1.668 & 2.542 & 13.437 & 32.847 & 3.899 & 1.269 & 11.498 & 1.806 &12.248 & 1.703 \\
002568654 & 4665 & 1.806 & 2.316 & 13.176 & 14.589 & 2.033 & 1.436 & 18.496 & 2.166 &19.073 & 2.029 \\
002583658 & 4863 & 1.762 & 2.444 & 12.575 & 36.688 & 4.217 & 1.313 & 11.037 & 1.790 &11.232 & 1.618 \\
002968820 & 4825 & 1.830 & 2.360 & 13.208 & 31.863 & 3.940 & 1.116 & 10.938 & 1.768 &11.333 & 1.681 \\
002970244 & 4901 & 2.009 & 2.239 & 12.515 & 48.662 & 5.472 & 1.093 &  8.729 & 1.599 & 8.193 & 1.527 \\
002984406 & 4812 & 2.083 & 2.117 & 12.841 & 35.259 & 4.096 & 1.289 & 11.184 & 1.783 &11.845 & 1.700 \\
003096721 & 4796 & 1.827 & 2.355 & 12.851 & 19.934 & 2.641 & 1.341 & 15.184 & 2.043 &14.468 & 1.890 \\
003112645 & 4781 & 1.424 & 2.757 & 13.566 & 30.938 & 3.883 & 1.068 & 10.885 & 1.748 &13.881 & 1.810 \\
003220783 & 4779 & 1.614 & 2.579 & 13.556 & 29.238 & 3.571 & 1.259 & 12.160 & 1.843 &16.802 & 2.032 \\
003340584 & 4788 & 2.024 & 2.177 & 11.281 & 16.093 & 2.171 & 1.541 & 18.126 & 2.193 &15.455 & 1.948 \\
003427365 & 5050 & 1.838 & 2.475 & 13.565 & 30.357 & 3.836 & 1.150 & 11.247 & 1.871 &10.009 & 1.744 \\
003525951 & 4856 & 1.710 & 2.474 & 12.926 & 35.374 & 4.295 & 1.091 & 10.252 & 1.723 &11.251 & 1.764 \\
003831992 & 5064 & 0.813 & 3.469 & 10.761 &202.042 &14.887 & 1.501 &  4.977 & 1.168 & 6.229 & 1.326 \\
003965419 & 4695 & 1.945 & 2.240 & 12.192 & 45.144 & 5.068 & 1.113 &  9.239 & 1.574 & 9.128 & 1.442 \\
005025172 & 4988 & 0.768 & 3.479 & 12.155 &239.682 &19.281 & 0.870 &  3.493 & 0.834 & 3.870 & 0.863 \\
009154402 & 5034 & 1.056 & 3.295 & 11.940 &209.337 &15.554 & 1.388 &  4.710 & 1.110 & 4.381 & 1.004 \\
010597307 & 5208 & 2.143 & 2.221 & 10.566 & 57.981 & 5.484 & 2.009 & 10.674 & 1.880 &11.948 & 1.854 \\
010964223 & 5175 & 2.052 & 2.299 & 10.341 & 30.240 & 3.975 & 1.022 & 10.562 & 1.859 &10.742 & 1.792 \\
011650806 & 5286 & 0.940 & 3.431 & 12.393 &219.136 &16.824 & 1.252 &  4.318 & 1.119 & 4.977 & 1.151 \\
011716190 & 5137 & 2.136 & 2.228 & 10.933 & 52.719 & 4.959 & 2.212 & 11.788 & 1.942 &11.014 & 1.797 \\
\hline
\end{tabular}
\end{table*}
Following the approach by \citet{Kallinger2010b} we use the values of $\Delta\nu$, $\nu_{\rm max}$ and the effective temperatures from the {\it Kepler} Input Catalog in combination with the scaling relations as described by \citet{Brown1991}, \citet{Kjeldsen1995}, and \citet{chaplin2008} to compute the masses and radii of the stars directly:
\begin{equation}
\nu_{\rm max} \approx \nu_{{\rm max}\odot}\frac{M/M_\odot}{(R/R_\odot)^2
\sqrt{T_{\rm eff}/T_{{\rm eff}\odot}}},
\end{equation}
\begin{equation}
\Delta\nu \approx \Delta\nu_\odot\sqrt{\frac{M/M_\odot}{(R/R_\odot)^3}}
\end{equation}
where $\nu_{{\rm max}\odot} = 3120$\,$\mu$Hz and  $\Delta\nu_\odot = 134.88$\,$\mu$Hz are solar values as determined by \citet{Kallinger2010a}. From the radii and \teff\ we computed the luminosity as $L$~$\propto$~$R^2 T_{\rm eff}^4$. 

$\nu_{\rm max}$ values were derived by fitting a Gaussian to the highest peaks in the periodogram at an appropriate frequency range of the comb-like structure, after removing a linear fit to the background noise level. For deriving $\Delta\nu$, we used the autocorrelation function of the periodogram over the range of the oscillations \citep{Huber2009}. The frequency spacing between consecutive {\it l} = 0 and 1 modes also gives rise to peaks in the autocorrelation function at $\frac{1}{2}\Delta\nu$. We selected those peaks closest to the value given by 
\begin{equation}
 \Delta\nu \approx a\nu_{\rm max}^b
 \label{eq:scalednu}
 \end{equation}
with  $a = 0.266 \pm 0.004$ and $b = 0.761 \pm 0.004$.

\begin{figure*}
\includegraphics[height=10cm,width=16cm]{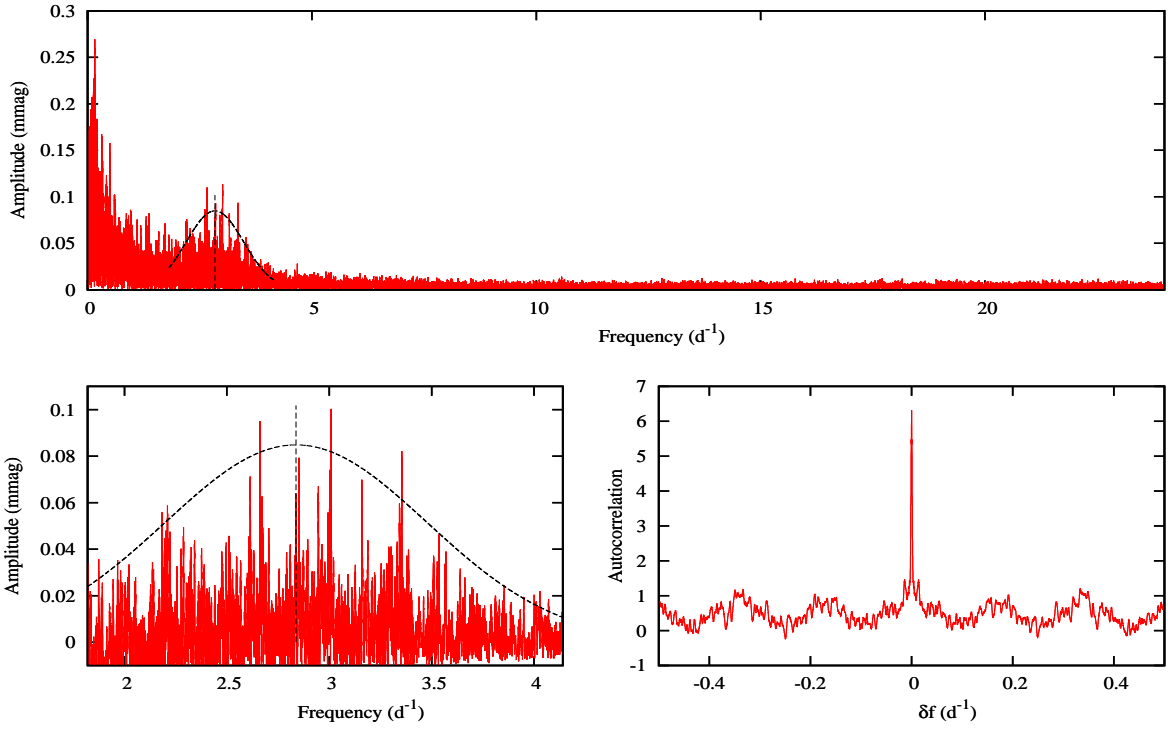}
\caption{Top panel: Periodogram of KIC 002568575 showing solar-like oscillations. Bottom panel left: expanded view of relevant region, right: autocorrelation function for 1.80 $<$ f $<$ 4.14~{\cycperday}.}
\label{autocorrelation}
\end{figure*}

\begin{figure*}
\includegraphics[height=10cm,width=16cm]{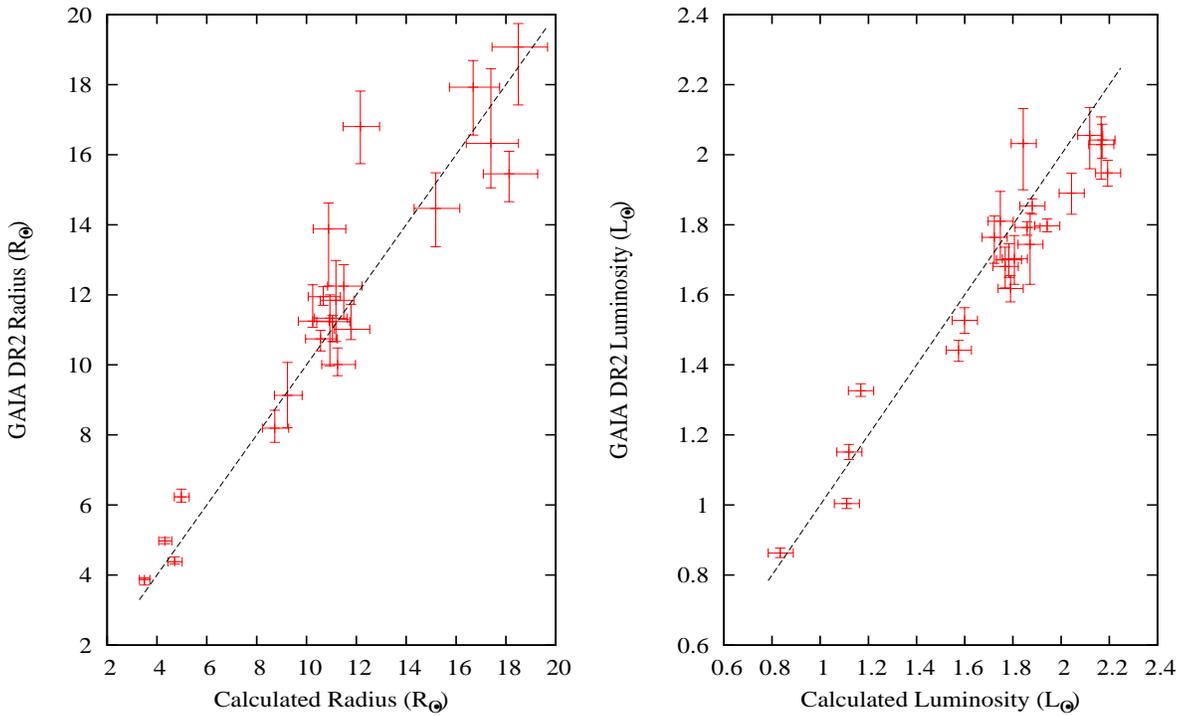}
\caption{Comparison of radii and luminosities derived from solar-like oscillations, with the radius and luminosities from GAIA DR2. A 1:1 ratio line is superimposed as a dashed line. The error bars in GAIA DR2 values are taken from \citet{Andrae2018} and the details of how our error values are estimated are discussed in the text.}
\label{gaia}
\end{figure*}

This empirical relationship is based on {\it Kepler} data of 662 field red giants \citep{Hekker2011}. As an example of our analysis, the periodogram of KIC\,002568575, together with the fitted Gaussian envelope and the autocorrelation function, is shown in Fig.\,\ref{autocorrelation}. The values of $\nu_{\rm max}$ and $\Delta\nu$ derived from our fits were used to calculate mass, radii and, consequently, luminosities, for the 23 newly-identified red giants for which we detected solar-like oscillations. Their details and calculated parameters are given in Table \ref{Sol_Like}. We have also shown the radius and luminosity from the GAIA Data Release 2 (DR2) \citep{Andrae2018} in the same table. We compared our $\nu_{\rm max}$ and $\Delta\nu$ values with \citet{Yu2018} for the common objects, and calculated the percentage difference for every common object by dividing the difference with our estimated values. We then took a standard deviation of these percentages differences and used it to find the propagation errors in radius and luminosity. The radius and luminosity, along with their roughly estimated errors are then plotted against GAIA DR2 values, as shown in Fig.\,\ref{gaia}. There is no significant difference between our estimated radius and luminosity from the solar-like oscillations and GAIA DR2 radius and luminosity. The average difference between our calculated and GAIA DR2 values, is 5\% in radius and 3\% in luminosity.

\begin{figure*}
\centering
\includegraphics[height=8cm,width=16cm]{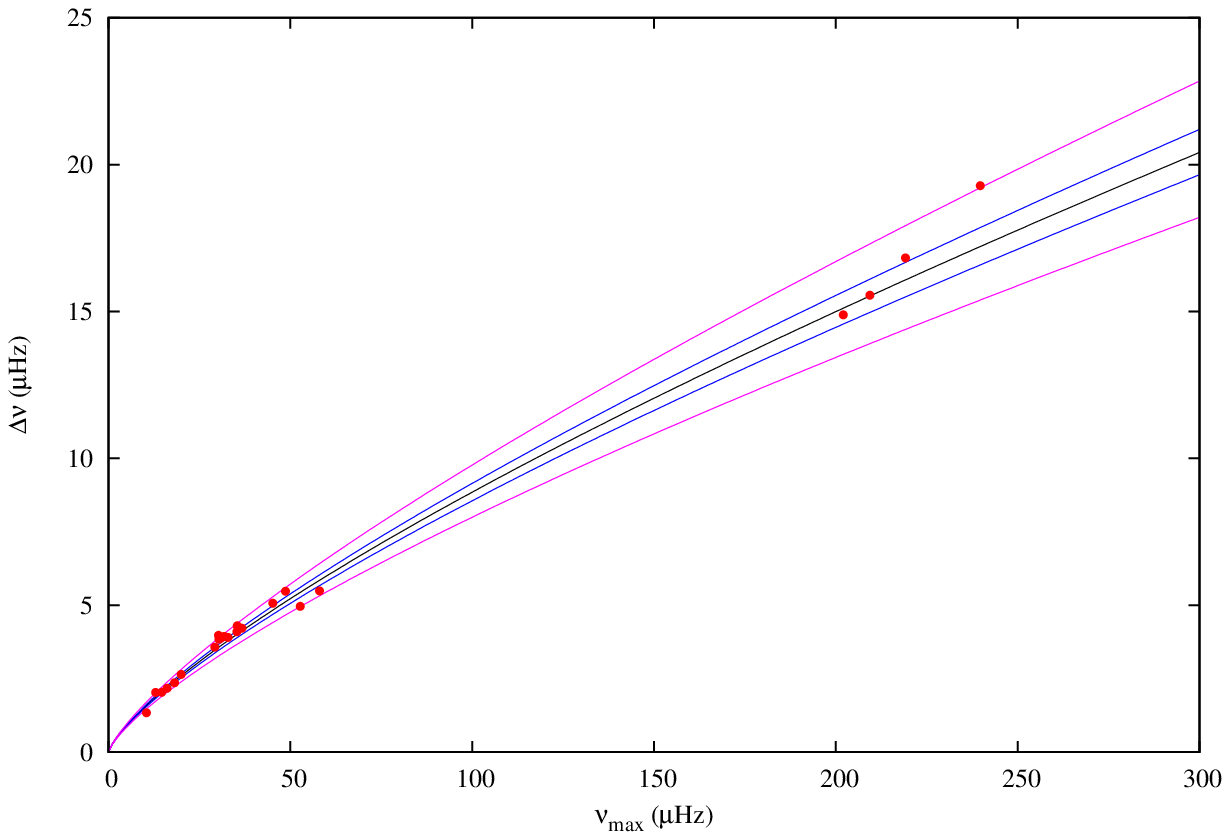}
\caption{The relationship between $\nu_{\rm max}$ and $\Delta\nu$ for stars given in Table \ref{Sol_Like} overplotted with the function $\Delta\nu \approx 0.266\nu_{\rm max}^{0.761}$ (solid black line), with the 1 sigma errors included in blue and 3 sigma errors included in magenta.}
\label{DeltanuvsNumax}
\end{figure*}

Fig.\,\ref{DeltanuvsNumax} shows the relationship between $\nu_{\rm max}$ and $\Delta\nu$ for our stars showing solar-like oscillations and how well they fit the known empirical relation given in Eq. (\ref{eq:scalednu}). The increased spread around the correlation at the high-end of the $\nu_{\rm max}$ relation is consistent with the expected spread,
which increases with $\nu_{\rm max}$ (see e.g. \citealt{Hekker2011,stello2009a,Huber2010}). Since the large separation, frequency of maximum amplitude, luminosity and the effective temperature are related to each other, one can use stellar evolution and pulsation models to constrain the solution. The stellar parameters are usually estimated by searching among a grid of stellar models to get a best fit for observed values of $\Delta\nu$, $\nu_{\rm max}$, \teff\ and metallicity. This is usually referred to as `grid' asteroseismology (\citealp{Stello2009b, Basu2010, Quirion2010, Gai2011}).

In our study, we have used the direct method to estimate the stellar parameters. This so-called direct method 
provides estimates that are independent of stellar evolutionary theory. The luminosities from the KIC and asteroseismic luminosity \logL$_{\nu}$ derived using scaling relations are not in good agreement because the estimated luminosity from the scaling relations is highly dependent on the accuracy of \teff. A typical error of 150\,K in the determination of \teff\ leads to an error of about 0.2 mag in the bolometric absolute magnitude. Therefore, spectroscopic observations are essential for the accurate determination of \teff\ and \logg\ \citep{Catanzaro2011, Niemczura2015}. The error in mass, radius and luminosity also depends on the error in estimating $\nu_{\rm max}$ and $\Delta\nu$. We manually selected the fits for each star that describes the overall shape of the envelope very well. It is difficult to estimate the uncertainties in $\nu_{\rm max}$ and $\Delta\nu$. \citet{Hekker2012} have investigated the uncertainties in these two quantities by means of simulations using different methods and found that for a time series of duration about 500 d, the uncertainty in $\nu_{\rm max}$ is about 2 \% and in $\Delta\nu$ it is about 1 \%. 

\begin{table*}[t]
\caption{Classification of non-radial pulsators in the nominal {\it Kepler} and {\it Kepler} K2 fields.}
\label{tab:classification}
\begin{center}
\centering
\begin{tabular}{cccccccc}
\hline
\hline
Field & $\delta$\,Sct & $\delta$\,Sct + Binary & $\gamma$\,Dor & $\gamma$\,Dor + Binary & Hybrids & Hybrid + Binary & Reference  \\
\hline
{\it Kepler}  & 403 & - & 441 & - & - & - & \citet{Debosscher2011} \\
{\it Kepler}  & 206 & 11 & 100 & 10 & 171 & 1 & \citet{Uytterhoeven2011} \\
{\it Kepler}  & 287 & - & 9 & - & - & - & \citet{Balona2013c}  \\
{\it Kepler}  & 84 & - & 207 & - & 32 & - & \citet{Bradley2015}  \\
{\it Kepler}  &983 & - &  -  & - &  - & - & \citet{Bowman2016}   \\
K2 & 377 & - & 133 & - & - & - & \citet{Armstrong2016}  \\
\hline
\end{tabular}
\end{center}
\end{table*}

Comparative studies between the direct method and independently determined properties from binaries, shows that the estimated parameters are found to be consistent at the level of precision of the uncertainties, i.e. up to 10\% or better \citep{Bruntt2010, Miglio2012b}. \citet{Silva2012} used the brightest solar-like oscillators from the {\it Kepler} data, and found excellent agreement between scaling relation inferred radius and those inferred from using Hipparcos parallaxes (at the level of a few percent). \citet{Huber2012} used combined Hipparcos parallaxes and interferometric observations of some of the brightest {\it Kepler} and CoRoT targets and also found excellent agreement with the stellar radii inferred from scaling relations, at the 5\% level. \citet{Gaulme2016} compared the masses and radii of 10 red giants obtained by combining the radial velocities and eclipse photometry with the estimates from the asteroseismic scaling relations, and found that the asteroseismic scaling relations overestimate red-giant radii by about 5 \% on average and masses by about 15 \% for stars at various stages of red-giant evolution.

\section{$\delta$\,Scuti, $\gamma$\,Doradus and hybrid stars} \label{sec:nonrad}

A detailed analysis of stellar pulsation frequencies can be used to infer interior stellar structure and test theoretical models. The main-sequence $\gamma$\,Dor and $\delta$\,Sct stars are particularly useful for asteroseismic studies. The $\delta$\,Sct stars pulsate in both radial and non-radial \pmodes\ 
with periods ranging from 18 min to 7 hrs (see \citealt{Breger2000} and \citealt{Bowman2018}, for a review). The pulsations in the $\delta$\,Sct variables are driven by the $\kappa$-mechanism operating in the HeII partial ionization zone with some additional contribution from the HI ionization zone. The $\gamma$\,Dor variables are generally cooler than $\delta$\,Sct stars and are mostly located near the cool edge of the $\delta$\,Sct instability strip. The pulsations in $\gamma$\,Dor variables are characterized by high-order, low-degree and multiple non-radial \gmodes\ with periods of 0.3 to 3 days \citep{Kaye1999, Balona2011a}. The pulsations in these stars are excited by the convective blocking mechanism at the base of their envelope convection zone \citep{Guzik2000}. The distinction between the two classes is clearer if we consider the value of the pulsation constant, Q, where all $\gamma$\,Dor stars have $Q$ $>$ 0.09 d and all $\delta$\,Sct stars have $Q$ $<$ 0.09 (see Fig. 9 of \citet{Handler2002}, for details), accommodating possible uncertainties in the determination of $Q$ \citep{Breger1990} Note that it is not always obvious to distinguish $\gamma$\,Dor variables from spotted stars as the expected pulsation and rotation periods are of the same order.

Due to the overlapping of \teff\ and \logg\ ranges for these two types of variables, there is a possibility for stars showing both types of pulsation simultaneously, i.e. $\delta$\,Sct-type combined with $\gamma$\,Dor-type pulsations, and are thus known as the A/F-type hybrid stars. Before the launch of the {\it Kepler} and CoRoT space missions, only a few stars pulsating in both {\it p} and \gmodes\ were  discovered and interpreted using theoretical models \citep{Bouabid2009}. Ground-based observations discovered only 4 hybrid stars \citep{Henry2005, Uytterhoeven2008, Handler2009}. The first published analysis on 234 targets by the {\it Kepler} Asteroseismic Science Consortium (KASC) confirmed the pulsations of either type ($\delta$\,Sct or $\gamma$\,Dor) and revealed the hybrid behavior in  essentially all of them \citep{2010ApJ...713L.192G}.  In a study of 750 A-F stars observed for four quarters, 475 stars showed either $\delta$\,Sct or $\gamma$\,Dor variability, and 36\% of these were hybrids \citep{Uytterhoeven2011}. 

\begin{sidewaystable*}
\centering
\caption{Newly identified 168 $\delta$\,Sct, 110 $\gamma$\,Dor and 72 $\gamma$\,Dor/$\delta$\,Sct hybrid variables in our study. The complete table is available online. ppt is parts per thousand.}
\label{Non_radial_puls_table}
\begin{tabular}{cccccccccccc}
\hline
\multicolumn{1}{c}{KIC} &
\multicolumn{1}{c}{Class} & 
\multicolumn{1}{c}{$\log T_{\rm eff}$} & 
\multicolumn{1}{c}{$\log (L/L_{\odot})$} & 
\multicolumn{1}{c}{\logg} & 
\multicolumn{1}{c}{Kp} & 
\multicolumn{1}{c}{$F_{puls}$} &  
\multicolumn{1}{c}{$P_{puls}$} & 
\multicolumn{1}{c}{$A_{puls}$} & 
\multicolumn{1}{c}{\logen} & 
\multicolumn{1}{c}{\logeff} &  
\multicolumn{1}{c}{\logamp} 
\\  
\multicolumn{1}{c}{(Id)} & 
\multicolumn{1}{c}{} & 
\multicolumn{1}{c}{(K)} & 
\multicolumn{1}{c}{ } & 
\multicolumn{1}{c}{$(cm s^{-2})$} & 
\multicolumn{1}{c}{(mag)} & 
\multicolumn{1}{c}{{$(\cycperday)$}} & 
\multicolumn{1}{c}{(days)} & 
\multicolumn{1}{c}{(ppt)} & 
\multicolumn{1}{c}{$({ppm^2}{\cpd})$} & 
\multicolumn{1}{c}{$(K^{-2}cm^{-2/3}s^{4/3})$} & 
\multicolumn{1}{c}{(ppm)}  \\ 
\hline
001433399 & GDOR & 3.836 & 0.786 & 4.055 & 12.84 &  0.579 & 1.727 &  0.296 &  4.467 & -8.076 & 2.471 \\
001575977 & GDOR & 3.850 & 0.783 & 4.120 & 13.61 &  2.002 & 0.499 &  1.277 &  6.815 & -8.110 & 3.106 \\
001872262 & GDOR & 3.834 & 0.547 & 4.266 & 13.74 &  1.989 & 0.502 &  0.527 &  6.040 & -8.089 & 2.721 \\
002162904 & DSCT & 3.853 & 1.322 & 3.652 & 12.66 & 14.290 & 0.070 &  2.016 &  8.919 & -8.080 & 3.304 \\
002282763 & GDOR & 3.891 & 1.329 & 3.822 & 12.46 &  0.949 & 1.053 &  0.447 &  5.255 & -8.171 & 2.650 \\
002283124 & DSCT & 3.834 & 0.582 & 4.232 & 12.62 & 14.585 & 0.068 &  0.091 &  6.248 & -8.085 & 1.960 \\
002304566 & GDOR & 3.831 & 0.684 & 4.130 & 13.18 &  2.900 & 0.344 &  0.106 &  4.977 & -8.073 & 2.026 \\
002425057 & GDOR & 3.884 & 1.687 & 3.477 & 12.60 &  3.388 & 0.295 &  0.064 &  4.672 & -8.130 & 1.806 \\
002581674 & GDOR & 3.852 & 1.460 & 3.532 & 11.28 &  1.275 & 0.784 &  0.108 &  4.284 & -8.070 & 2.036 \\
002583748 & GDOR & 3.826 & 0.783 & 4.021 & 13.64 &  2.443 & 0.409 &  1.047 &  6.816 & -8.056 & 3.020 \\
002693450 & GDDS & 3.848 & 0.836 & 4.061 & 12.71 &  2.254 & 0.443 &  0.099 &  4.701 & -8.101 & 1.997 \\
002837174 & DSCT & 3.855 & 0.880 & 4.053 & 11.78 & 15.263 & 0.065 &  0.117 &  6.504 & -8.115 & 2.068 \\
002988783 & GDOR & 3.854 & 0.863 & 4.067 & 13.75 &  2.719 & 0.367 &  0.548 &  6.347 & -8.115 & 2.739 \\
003230227 & GDDS & 3.901 & 1.297 & 3.893 &  9.00 &  0.141 & 7.052 &  0.652 &  3.932 & -8.196 & 2.814 \\
003240550 & GDDS & 3.817 & 0.512 & 4.226 & 12.24 &  1.028 & 0.972 &  2.960 &  6.966 & -8.052 & 3.471 \\
003245774 & GDOR & 3.818 & 0.411 & 4.322 & 11.96 &  0.528 & 1.892 &  0.051 &  2.868 & -8.059 & 1.710 \\
003340360 & GDDS & 3.863 & 0.720 & 4.234 & 13.84 &  3.008 & 0.332 &  0.279 &  5.848 & -8.144 & 2.446 \\
003356155 & DSCT & 3.859 & 0.838 & 4.111 & 12.82 & 16.091 & 0.062 &  0.619 &  7.997 & -8.128 & 2.792 \\
003441864 & DSCT & 3.923 & 1.113 & 4.166 & 12.37 & 15.131 & 0.066 &  0.147 &  6.698 & -8.260 & 2.169 \\
003445406 & GDOR & 3.842 & 0.639 & 4.216 & 13.15 &  2.412 & 0.414 &  0.573 &  6.281 & -8.101 & 2.758 \\
\hline
\end{tabular}
\end{sidewaystable*}

\begin{figure*}[t]
\includegraphics[height=10cm,width=16cm]{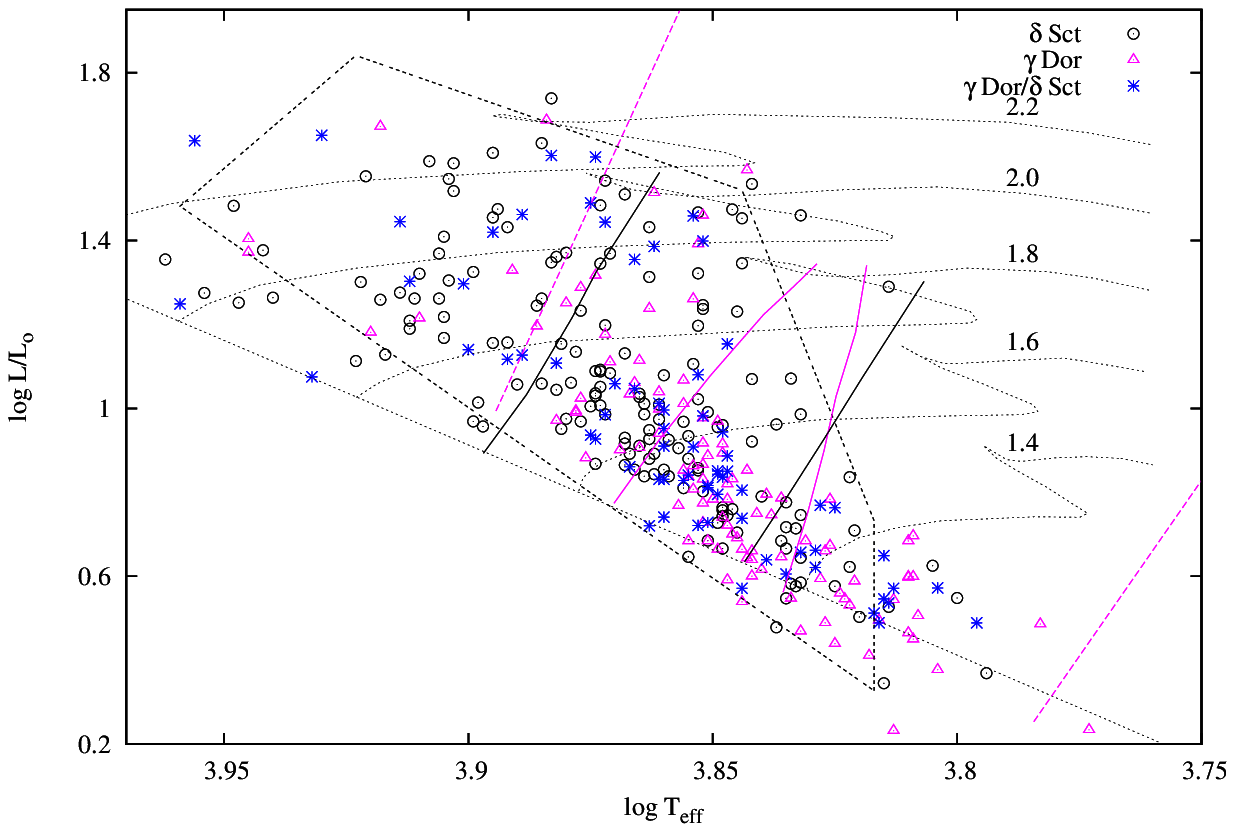}
\caption{Location of newly discovered candidate $\delta$\,Sct, $\gamma$\,Dor and $\gamma$\,Dor/$\delta$\,Sct hybrid variables in the theoretical HR diagram. The open black circles are $\delta$\,Sct stars and the trapezoidal region is the general location of {\it Kepler} $\delta$\,Sct stars \citep{Balona2011b}. The $\gamma$\,Dor stars and the hybrids are shown in open magenta triangles and blue stars, respectively. Also shown is the zero-age main sequence and evolutionary tracks for models with 1.4, 1.6, 1.8, 2.0 and 2.2 {$\msun$}. 
The $\delta$\,Sct red and blue edges for the p1 radial mode, calculated by \citet{Dupret2004} are shown in black solid lines. 
The $\gamma$\,Dor red and blue edges for {\it l} = 1, calculated by \citet{Dupret2004} are shown in magenta solid lines. The red and blue edges of the $\gamma$\,Dor instability strip calculated by \citet{Xiong2016} are shown in the magenta dashed lines.}
\label{HR_DIAGRAM}
\end{figure*}

However, \citet{Balona2014} found that essentially all $\delta$\,Sct stars in the {\it Kepler} field contain low-frequency modes typically associated with \gmodes\ and concluded that all $\delta$\,Sct stars should therefore be considered as `hybrids'. 
He also concluded that the presence of $\delta$\,Sct pulsations is strongly connected with the low frequencies in $\delta$\,Sct stars.
In accordance with this finding, it is clear that so-called $\delta$\,Sct/$\gamma$\,Dor hybrids (according to \citealt{2010ApJ...713L.192G}) actually occur throughout the $\delta$\,Sct instability strip. 
This supports Balona's assertion that the $\delta$\,Sct/$\gamma$\,Dor hybrid classification is redundant. In this paper, we classify $\delta$\,Sct/$\gamma$\,Dor stars as $\delta$\,Sct variables and we still refer to stars with their dominant frequencies in the low ({\it g-}mode) range but also displaying high-frequency modes as $\gamma$\,Dor/$\delta$\,Sct hybrids. Hybrid stars are among the most interesting and vital targets for asteroseismology for providing additional constraints on stellar structure, because the  $\gamma$\,Dor stars pulsate in \gmodes\ which have high amplitudes deep in the star and allow us to probe the stellar core, while the \pmodes, efficient in $\delta$\,Sct stars, have high amplitudes in the outer regions of the star and probe the stellar envelope. Using the {\it Kepler} archive time-series data, \citet{Kurtz2014} determined the rotation rate in the deep core and surface of the main sequence hybrid star KIC\, 11145123 and found that the surface rotates slightly faster than the core.  However, our understanding of the structure of the outer layers, the convective core and the outer convective zone is still far from complete, hence a better understanding of the driving mechanisms in these stars is very important.


\begin{table*}[t]
\caption{Peak values in the distribution of $Amplitude$ (\logamp), $Energy$ (\logen) and $Efficiency$  (\logeff), for $\delta$\,Sct, $\gamma$\,Dor/$\delta$\,Sct hybrids and $\gamma$\,Dor variables, along with their standard deviations.}
\label{Ene_Eff_Dist_Peak_tab}
\begin{center}
\centering
\begin{tabular}{lcccccc}
\hline
\hline
Class & \logamp & $\sigma$ & \logen & $\sigma$ & \logeff & $\sigma$  \\
  & (ppm) & (ppm) & $(ppm^2\cpd)$  & $(ppm^2\cpd)$  & $(K^{-2}cm^{-2/3}s^{4/3})$  & $(K^{-2}cm^{-2/3}s^{4/3})$ \\
\hline
$\delta$\,Sct                 & 2.668 & 0.801 & 7.812 & 1.430 & -8.117 & 0.047 \\
$\gamma$\,Dor/$\delta$\,Sct & 2.442 & 0.572 & 5.436 & 1.377 & -8.102 & 0.036 \\
$\gamma$\,Dor                & 2.554 & 0.627 & 5.548 & 1.568 & -8.098 & 0.034\\
\hline
\end{tabular}
\end{center}
\end{table*}

\begin{figure*}[t]
\centering
\includegraphics[height=10cm,width=16cm]{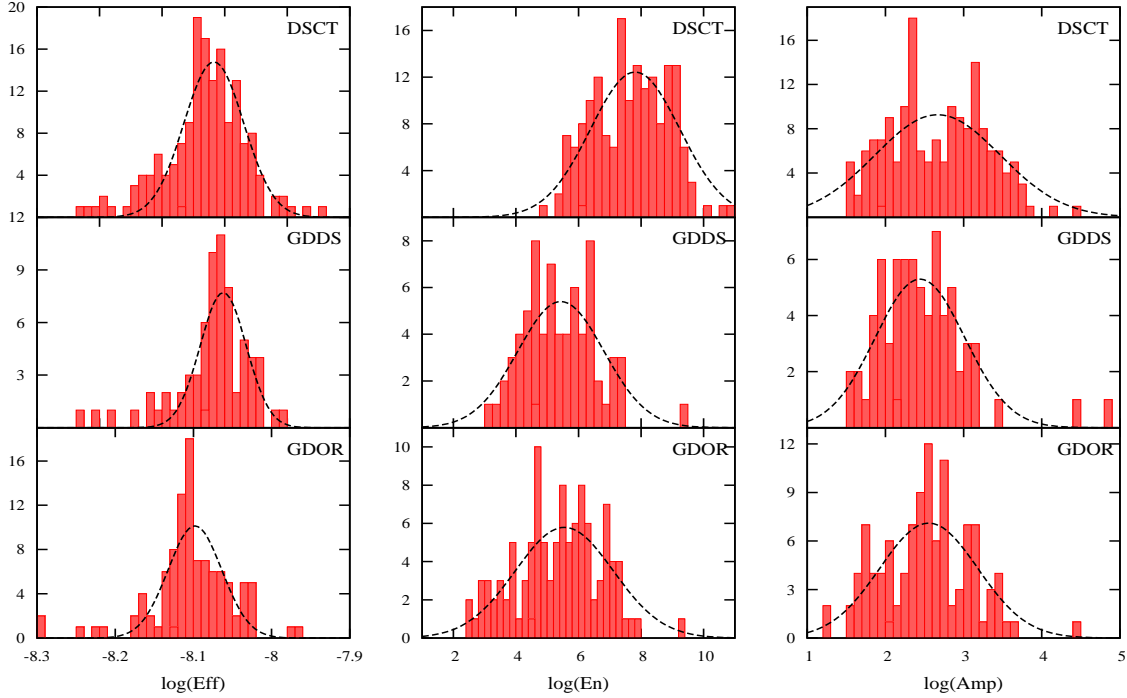}
\caption{
Distribution of $Efficiency$ (\logeff), $Energy$ (\logen) and $Amplitude$ (\logamp). The first and third row shows the distribution for $\delta$\,Sct and $\gamma$\,Dor variables, respectively. Distribution for $\gamma$\,Dor/$\delta$\,Sct hybrids are shown in the second row. The Y-axis represents the number of stars belonging to each bin(N). The dashed black line is the gaussian fit to each of the histogram, for the peak estimation.
}
\label{Log(eff)-Log(ene)-Log(amp)Dist}
\end{figure*}

\begin{figure*}
\centering
\includegraphics[height=9cm,width=14cm]{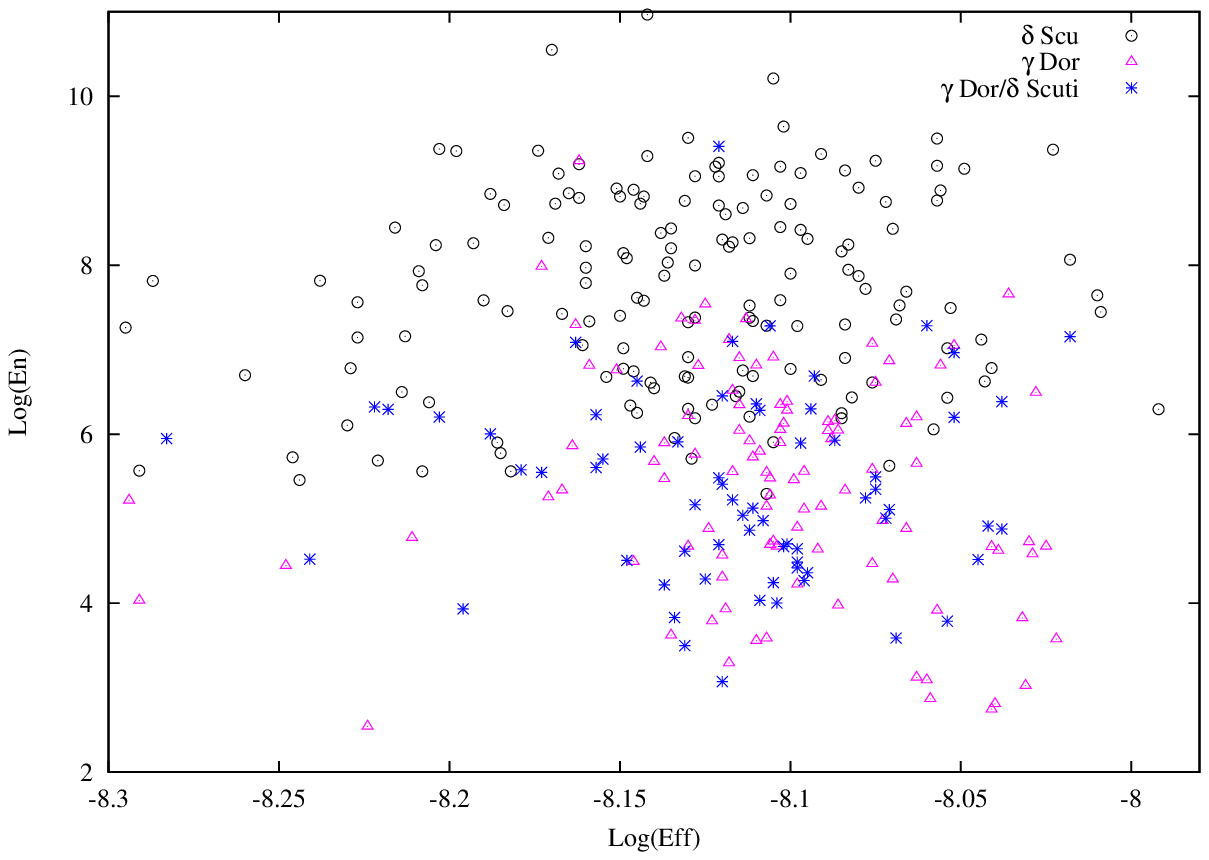}
\caption{The \logeff-\logen\ diagram for the stars characterized as $\delta$\,Sct and $\gamma$\,Dor and $\gamma$\,Dor/$\delta$\,Sct hybrids. The $\delta$\,Sct, $\gamma$\,Dor and the hybrid variables are represented in open black circles, open magenta triangles and blue stars respectively. }
\label{Log(eff)-Log(ene)_diag}
\end{figure*}

Some major previous efforts to classify the non-radial pulsators in the {\it Kepler} and K2 fields include those mentioned in Table\,\ref{tab:classification}. Combining all these studies results in around 1893 candidate non-radial pulsators in the nominal {\it Kepler} field (counting the overlapping objects once). This is an approximation, since it includes the 403 candidate $\delta$\,Sct/$\beta$\,Cephei ($\beta$\,Cep) stars and 441 candidate $\gamma$\,Dor/slowly pulsating B (SPB) stars given by \citet{Debosscher2011}, who could not reliably distinguish between $\delta$\,Sct and $\beta$\,Cep stars, nor between $\gamma$\,Dor and SPB stars, because their classifier only used information from the {\it Kepler} light curves. Moreover, since there are no other previous major studies dedicated to the classification of non-radial pulsators in the {\it Kepler} field, we are certain that the error in our approximation is small. Very recently, \citet{Ibanoglu2018} compiled a comprehensive catalog of $\gamma$\,Dor pulsators, listing 109 genuine variables, 291 candidate stars in the CoRoT fields, 318 candidate variables in {\it Kepler} field including 11 objects that are in a binary system. They have included one star which was classified as binary+hybrid by \citet{Uytterhoeven2011}, in their list of 11 candidate binary+$\gamma$\,Dor variables. In addition, they also compiled a list of 233 hybrids of which 205 are in the {\it Kepler} field.

We identified the candidate $\delta$\,Sct and $\gamma$\,Dor variables from a large sample of {\it Kepler} field stars, using automated constraints (frequency and amplitude), followed by a visual inspection of all the light curves and the frequency spectra (DFT). In addition to frequency, the pulsation amplitudes were also used to potentially filter out the high degree modes that generally would be expected to have a lower photometric amplitude. We also used the AoV algorithm in multi-harmonic mode \citep{Czerny1989, Schwarzenberg1996} to estimate and confirm the periods of these stars, but list the periods derived from the DFT in Table \ref{Non_radial_puls_table}. We characterized the stars as $\delta$\,Sct if most of the detected frequencies in the DFT are between 5~{\cycperday} and  24~{\cycperday} (being the Nyquist frequency for LC {\it Kepler} data). Stars where the lower frequencies (less than 5~{\cycperday}) are also present, are also classified as $\delta$\,Sct variables. Stars for which the detected frequencies were between 0.30 and 5~{\cycperday} are characterized as $\gamma$\,Dor variables. By carefully examining the periodograms of $\delta$\,Sct stars and $\gamma$\,Dor stars we were also able to identify the $\gamma$\,Dor/$\delta$\,Sct hybrids.

This search led to the discovery of 168, 110, and 72 new candidate $\delta$\,Sct stars, $\gamma$\,Dor stars, and $\gamma$\,Dor/$\delta$\,Sct variables, respectively. They are listed in Table \ref{Non_radial_puls_table} and their positions in the H-R diagram are shown in Fig.\,\ref{HR_DIAGRAM}.
Since we included the $\delta$\,Sct/$\gamma$\,Dor variables in the $\delta$\,Sct variable definition, we do not compare our number of candidate $\delta$\,Sct variables to those mentioned in the literature. Comparing the number of candidate non-radial pulsators detected in our study with the literature shows that, our study increases the number of non-radial pulsators in the {\it Kepler} field, by $\sim$20\% (counting the overlapping objects only once). In comparison with the catalogue of \citet{Ibanoglu2018}, for $\gamma$\,Dor variables, our study increases the number of candidate $\gamma$\,Dor variables in the {\it Kepler} field by $\sim$34\%, and the overall candidate $\gamma$\,Dor variables by $\sim$15\% ($\sim$12\% after including the 133 candidate $\gamma$\,Dor pulsators detected by \citet{Armstrong2016}, which was not included in the catalogue of \citet{Ibanoglu2018}). Note that \citet{Ibanoglu2018} did not include the $\gamma$\,Dor/SPB variables from \citet{Debosscher2011} in their catalogue of $\gamma$\,Dor variables, so the percentage increase for $\gamma$\,Dor stars stated above does not include them. 

\citet{Bouabid2013} showed that some stars classified as a potential hybrid star from the frequency distribution could in fact be rapidly rotating $\gamma$\,Dor stars whose \gmodes\ have been shifted to higher frequencies. To confirm the classification of the hybrid stars and to determine the nature of their pulsations as {\it p-} or \gmodes, multi-colour photometry and/or a spectroscopic study is required for the determination of the atmospheric parameters \citep{Catanzaro2010}. The candidate $\delta$\,Sct, $\gamma$\,Dor and $\gamma$\,Dor/$\delta$\,Sct hybrids found in our study could be the potential targets for detailed ground-based photometry and mid- to high-resolution spectroscopic observations because the fundamental parameters (\teff, \logg, \vsini, and \logL) are essential for computing asteroseismic models and interpreting {\it Kepler} data. 


To understand the relationship between $\delta$\,Sct, $\gamma$\,Dor and the hybrid pulsators, \citet{Uytterhoeven2011} first introduced two new empirical observables, $Energy$ (related to the energy contained in the oscillation) and $Efficiency$ (related to the convective efficiency of the outer convective zone), using the available observational data:
\begin{equation}
Energy \equiv (A_{\rm max} \zeta_{\rm max})^2,
\end{equation}
\begin{equation}
{\it Efficiency}\equiv (T_{\rm eff}^3\log g)^{-2/3},
\end{equation}
where $A_{\rm max}$ and $\zeta_{\rm max}$ refer to the highest amplitude mode of the star (in ppm) and its frequency (in d$^{-1}$), respectively.
\citet{Uytterhoeven2011} found that the distribution in the logarithm of $Efficiency$ and $Energy$ peaks at different values for $\delta$\,Sct and $\gamma$\,Dor stars. 
We further examined the peak values of the distribution in the logarithm of $Efficiency$ (\logeff), $Energy$ (\logen) and $Amplitude$ (\logamp), for our large sample of non-radial pulsators, where $Amplitude$ here refers to the highest amplitude mode of the star in the DFT of the light curve. The results are shown in Table \ref{Ene_Eff_Dist_Peak_tab}. The values in the table are derived by fitting a gaussian to the histograms. The distributions of \logen, \logamp, and \logeff\ along with the fitted gaussians are shown in Fig.\,\ref{Log(eff)-Log(ene)-Log(amp)Dist}. The majority of $\delta$\,Sct stars have \logen\ $>$ 7.5 while, for the $\gamma$\,Dor variables, the majority of stars in the distributions almost all have \logen\ $<$ 7.5. 

The \logen\ versus \logeff\ diagram is plotted in Fig.\,\ref{Log(eff)-Log(ene)_diag}. Although not much can be inferred about these observables and its physical basis at this moment, it is certain that these observables peak at different values in their respective distributions (see Fig. \ref{Log(eff)-Log(ene)-Log(amp)Dist} and Table \ref{Ene_Eff_Dist_Peak_tab}). As a result, it can be used to some extent, as a method to distinguish the non-radial pulsators, especially using the distribution in the logarithm of $Energy$ (\logen). Our study confirmed and improved the previous results obtained by \citet{Uytterhoeven2011}, who used fewer non-radial pulsators in deriving the distribution peaks, based on their new observables.

\section{Rotational Variables} \label{sec:rotvar}

\begin{table*}[t]
\centering
\caption{513 newly identified stars with rotational modulation (starspots). All the missing data in the table are represented by -1.0. The complete table is available online.}
\label{Rot-var_table}
\begin{tabular}{ccccccccccc}
\hline
\multicolumn{1}{c}{KIC} & 
\multicolumn{1}{c}{$F_{rot}$} & 
\multicolumn{1}{c}{$P_{rot}$} &
\multicolumn{1}{c}{$A_{rot}$} & 
\multicolumn{1}{c}{\teff} & 
\multicolumn{1}{c}{\logT} & 
\multicolumn{1}{c}{$\log (L/$L$_{\odot})$} &
\multicolumn{1}{c}{\logg} &
\multicolumn{1}{c}{Radius} &  
\multicolumn{1}{c}{Kp} &  
\multicolumn{1}{c}{(B-V)$_0$}  \\  

\multicolumn{1}{c}{(Id)} & 
\multicolumn{1}{c}{{(\cycperday)}} & 
\multicolumn{1}{c}{(days)} & 
\multicolumn{1}{c}{(ppt)} & 
\multicolumn{1}{c}{(K)} & 
\multicolumn{1}{c}{(K)} & 
\multicolumn{1}{c}{ } & 
\multicolumn{1}{c}{$(cm s^{-2})$} & 
\multicolumn{1}{c}{$(\rsun)$} & 
\multicolumn{1}{c}{(mag)} & 
\multicolumn{1}{c}{(mag)}   \\ 
\hline
000757280 & 0.7406 & 1.3502 & 0.1620 & 6648 & 3.823 & 0.696 & 4.082 & 1.683 & 11.901 & 0.375 \\
001026133 & 0.3714 & 2.6927 & 0.1152 & 6825 & 3.834 & 0.650 & 4.171 & 1.515 & 13.161 & 0.331 \\
001431474 & 0.2536 & 3.9433 & 0.4474 & 6645 & 3.822 & 0.749 & 4.034 & 1.791 & 13.521 & 0.412 \\
001571732 & 0.4615 & 2.1670 & 0.3475 & 6507 & 3.813 & 0.672 & 4.065 & 1.710 & 12.872 & 0.395 \\
001575672 & 0.1121 & 8.9227 & 0.0882 & 6532 & 3.815 & 0.663 & 4.084 & 1.679 & 11.987 & 0.390 \\
002142575 & 0.6343 & 1.5765 & 0.0334 &-1.00 &-1.000 &-1.000 &-1.000 &-1.000 &  9.849 & 0.432 \\
002155343 & 0.1233 & 8.1093 & 0.1484 & 6662 & 3.824 & 0.843 & 3.953 & 1.986 & 13.307 & 0.381 \\
002297398 & 0.3154 & 3.1703 & 0.5252 & 4907 & 3.691 &-0.371 & 4.482 & 0.905 & 15.833 & 0.918 \\
002311218 & 0.1257 & 7.9527 & 0.1002 & 6424 & 3.808 & 0.400 & 4.288 & 1.282 & 12.159 & 0.323 \\
002436421 & 0.6954 & 1.4381 & 0.3304 & 6881 & 3.838 &-1.000 & 3.956 &-1.000 & 15.017 & 0.418 \\
002437762 & 0.2356 & 4.2439 & 0.7985 & 7489 & 3.874 & 1.106 & 3.939 & 2.126 & 15.014 & 0.335 \\
002437888 & 0.1954 & 5.1171 & 0.7699 & 6148 & 3.789 & 0.651 & 3.985 & 1.868 & 14.784 & 0.600 \\
002438566 & 0.1421 & 7.0374 & 0.9979 & 6577 & 3.818 & 0.773 & 3.993 & 1.880 & 15.932 & 0.453 \\
002578513 & 0.2743 & 3.6454 & 0.1243 & 6800 & 3.833 & 1.357 & 3.539 & 3.445 & 11.102 & 0.245 \\
002715342 & 0.1311 & 7.6297 & 0.0707 & 6256 & 3.796 & 0.444 & 4.199 & 1.423 & 12.178 & 0.365 \\
002715487 & 1.4276 & 0.7005 & 0.1594 & 7459 & 3.873 & 1.172 & 3.873 & 2.313 & 13.809 & 0.230 \\
002833554 & 0.1121 & 8.9211 & 0.1409 & 5486 & 3.739 & 0.485 & 3.941 & 1.939 & 12.162 & 0.597 \\
002852641 & 0.5296 & 1.8881 & 0.1124 & 6533 & 3.815 & 0.439 & 4.283 & 1.297 & 12.084 & 0.321 \\
002857477 & 0.6818 & 1.4666 & 0.1811 & 6741 & 3.829 & 0.679 & 4.123 & 1.605 & 12.969 & 0.310 \\
002860732 & 0.1282 & 7.8006 & 0.1103 & 6622 & 3.821 & 0.528 & 4.226 & 1.399 & 12.574 & 0.328 \\
\hline
\end{tabular}
\end{table*}

It was found a long time ago that the hot B stars are the most rapidly rotating main sequence stars and that rotation slows down for cooler stars. For stars cooler than F5, the rate of rotation drops below 10 {\kms} which is difficult to measure. The reason why rotation is so slow for cool stars like the Sun is thought to be due to convection. The stars cooler than F5 have outer convective envelopes, while hotter stars have radiative envelopes. Convection is thought to be necessary for the generation of a magnetic field by the dynamo mechanism \citep{Reiners2012}. If a star has a magnetic field with surface convective envelopes, that means the ionized gases (which are conducting) are trapped by the magnetic lines of flux. Ionized gas in the stellar wind is forced to move along the magnetic field lines, which carry away the angular momentum and hence there is a braking action on rotation. Another product of a magnetic field is that it can create starspots and flares. Therefore one only expects stars cooler than F5 to have starspots and to show flares. 

It was shown by \citet{Skumanich1972} that the rotation periods of solar-type stars decrease over time, $t$, such that the rotational velocity, $v_{\rm rot} \propto 1/\sqrt{t}$. In a similar way, the chromospheric activity, which is a proxy for magnetic field strength, also decreases over time. The cause of the Skumanich law is believed to be angular momentum loss due to stellar winds, but the exact dependence of $v_{\rm rot}$ on age is still not understood properly. However, this has not hindered the development and use of `gyrochronology', a technique to determine the ages of field stars which is based upon the period--age--mass (PtM) relationship \citep{Barnes2007, Mamajek2008, Kawaler1989, Barnes2010}. 

Using observed rotation periods, it was shown by \citet{Barnes2001}, that the age dependence of rotation for these stars is of Skumanich type ($P \propto \sqrt{t}$) and that the correlation of stellar mass (or, more specifically, its proxy \bvo), with surface rotation rate that is seen in the Hyades cluster, also extends to stars of types F5 and later in general. \citet{Barnes2003} demonstrated the consistency of this correlation in a detailed study including many open clusters, finding that the correlation becomes tighter with increasing cluster age. The PtM relationship is not unique for stars of young ($\leq$100 Myr) clusters, and two (fast and slow) branches emerge from the PtM surface \citep{Barnes2003}. The fast branch vanishes with increasing age and the stars come together in their rotational evolution onto the slow branch, leading to an unique relationship between P and M at an age of about 600 Myr.

The {\it Kepler} data are ideal for measuring the rotational period of spotted stars. Rotation can be measured at the surface of individual stars using either spectroscopy or periodic variations in photometric light curves due to the presence of these starspots \citep{Mosser2009}.  
Previously, stellar rotation was measured using spots, rotational broadening of absorption lines (e.g., \citealt{Kaler1989}) and chromospheric activity, specifically Ca II emission \citep{Noyes1984}. 
With the launch of the CoRoT and {\it Kepler} telescopes, the study of stellar rotation has seen significant advances and has reached a new level of understanding. The high precision data from these space missions allow to estimate the rotational period for a well-defined large sample of the slowly rotating stars with low-amplitude modulation. Some previous studies to derive the rotational periods focusing on a broader {\it Kepler} sample are those of:
\citet{Reinhold2013} (RRB13 in what follows), with an emphasis on differential rotation, who derive the rotation period of $\sim$24\,000 stars using data from {\it Q3}; \citet{Nielsen2013}, who measured the rotation period of $\sim$12\,000 stars with {\it Q2--Q9} data, comparing their results with previous spectroscopic studies; \citet{McQuillan2014}, who derived rotation periods for $\sim$34\,000 main-sequence stars of temperature less than 6\,500\,K.

\begin{figure*}
\includegraphics[height=10cm,width=16cm]{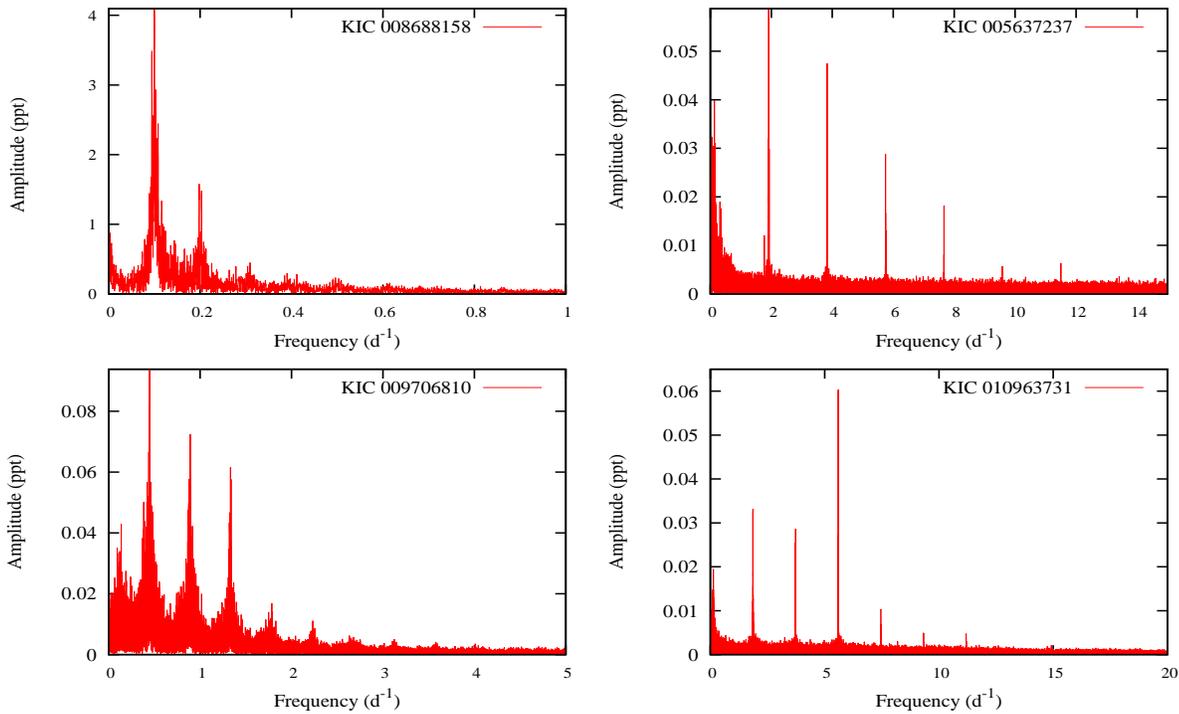}
\caption{Periodograms of some rotational variables, showing spots. The four panels show examples of harmonics for rotation frequencies covering the full range detected, i.e. from rotation frequencies of 0.1 \cycperday\ (upper left panel) to 6.0 \cycperday\ (lower right panel).}
\label{starspots}
\end{figure*}

\citet{Meibom2011}, who exploited the {\it Kepler} data to determine rotation periods in the cluster NGC\,6811, confirmed the existence of a unique functional relationship between rotation period, colour and age. However the exact form of this relationship is yet to be investigated. More recently, \citet{Reinhold2015} derived rotation and differential rotation periods for more than 18,500 and 12,300 stars respectively, using different approaches, thereby confirming that the relative shear 
$\alpha$ (a measure of differential rotation), increases with rotation period for stars with \teff 
$<$ 6700 K, while hotter stars show the opposite behaviour. From the mean rotation periods and their uncertainities, they further infer stellar ages between 100 Myr and 10Gyr for more than 17,000 stars using different gyrochronology relations. Gyrochronology is calibrated using derived rotation period and age for cool stars of different masses, for stars in clusters with known ages. Until recently, it was not possible to measure rotation periods and test the gyrochronology relation for stars older than about one billion years, so model predictions were used to infer gyrochronology ages. Now, the rotation periods are known for stars in an open cluster of intermediate age (NGC 6819, 2.5 Gyr old, \citet{Meibom2015}), and for old field stars with asteroseismologically determined ages. This confirmed the expected relationship between rotation period and stellar mass at the cluster age with a precision of order 10 \%, but failed to describe the asteroseismic sample \citep{Angus2015}. \citet{Saders2016} essentially confirmed the presence of unexpectedly rapid rotation in stars that are more evolved than the Sun and showed that after incorporating dramatically weakened magnetic braking in the stellar evolutionary models, for old stars, both the cluster and asteroseismic data can be explained. This weakened braking has a significant impact on the gyrochronology relation for stars, including our Sun, that are more than midway through their main-sequence lifetimes.

\begin{figure*}[t]
\centering
\includegraphics{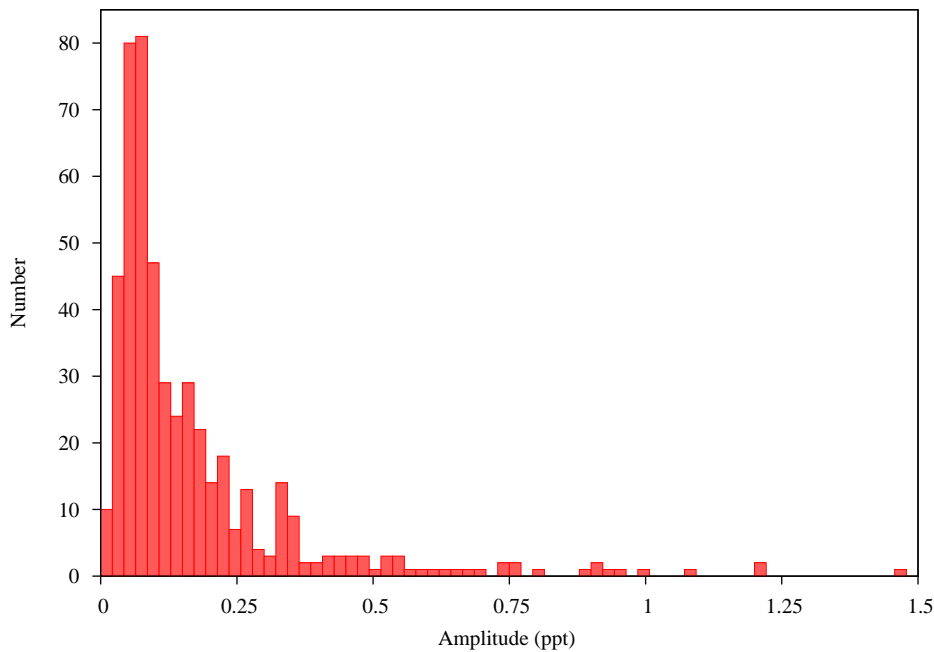}
\caption{Distribution of amplitudes of stars with a low frequency peak and harmonics. Only amplitudes smaller than 1.5 ppt are shown; the distribution has a long tail.}
\label{rotvar-NumvsAmphist}
\end{figure*}

We examined 15\,106 {\it Kepler} field stars showing light variations that could be attributed to starspots. This was done by examining the periodogram (DFT) and light curve of each star and assigning a variability type. The DFT is cross-matched with the AoV periodograms to make sure peaks are not spurious. Often the peaks are not single, but multiple. If the frequency ratios are equal to integer numbers, the extra peaks represent harmonics. Spots on a rotating stellar surface will exhibit harmonic frequencies because they act like low-level eclipses which consist of a flat light curve with dips. Since this is not a sinusoidal signal, harmonics will be present, depending on the spot location (its latitude), and the inclination angle.
Sometimes, peaks are multiple without harmonics, which probably means there are multiple spots migrating with respect to each other and/or changing in size. 
We took the presence of harmonics as a good indicator of starspots. In doubtful cases, we also checked the folded light curves of the stars, based on the derived rotation frequencies. In this way we can distinguish between pulsational variables of the $\gamma$\,Dor type and spotted stars. 

We measured the rotation periods of 513 stars where we assume that the variation is due to starspots and rotation. The periodograms of some of the rotational variables are shown in Fig.\,\ref{starspots}. 
There is a possibility of contamination of light curves by background stars. However, out of a sample of 513 stars 
, it is not statistically possible that more than a minute fraction of these stars could have their values of \bvo\ artificially raised by hotter (albeit more distant) background stars, for the following reasons: A background star will only rarely coincide closely enough with a target star to contaminate its light curve; background stars will be more distant and will have a small effect (if any) on a target star's light curve and the number density of stars falls off rapidly with temperature - so that the likelihood of contamination by a hotter star is small. It is also unlikely that SPB stars have been misidentified as rotational variables, since SPB stars are rare and do not tend to show the clear harmonic patterns displayed in Fig.\,\ref{starspots}. The possibility remains that cool background stars or close companions could be responsible for some of the detections of rotational modulation. We do not expect the incidence of such contamination to be a common occurrence in our sample, as cooler background stars would already provide a weaker total light signal than our candidate stars due to cooler surface temperature and greater distance, so that the very small effects (on top of the total light signal) that are due to rotational modulation (see Fig.\,\ref{rotvar-NumvsAmphist} in this paper) would not always show up as a measurable effect in our candidate stars' light curves. Still, it is possible that the light curves of some of our candidate stars with rotational modulation have been contaminated by distant stars in the same line of sight, or by close companions. We checked contamination flags in the MAST archive and found that most of our 513 candidates had contamination values less than 0.1 and only 29 objects had values between 0.1 and 0.55.


The newly discovered rotational variables are listed in Table \ref{Rot-var_table} (complete table is provided online). Fig.\,\ref{rotvar-NumvsAmphist} shows the distribution of amplitudes of the peak at the rotational frequency. It appears that the number of stars with starspots increases with decreasing amplitude. The number decreases below 0.03 ppt - that could be due to the difficulty in detecting peaks below this amplitude. There is also a very long tail towards the higher amplitudes.
In fact, about 4.4\% of stars have amplitudes larger than 1 ppt and there are 2 stars where amplitudes are higher than 10 ppt. 
Such large amplitudes are seen even in B type stars with spots, for example, HD131120 \citep{Briquet2001}.
\\
\\
\\
\subsection{The Period-Colour Relationship}
\begin{figure*}
\centering
\includegraphics[height=10cm,width=16cm]{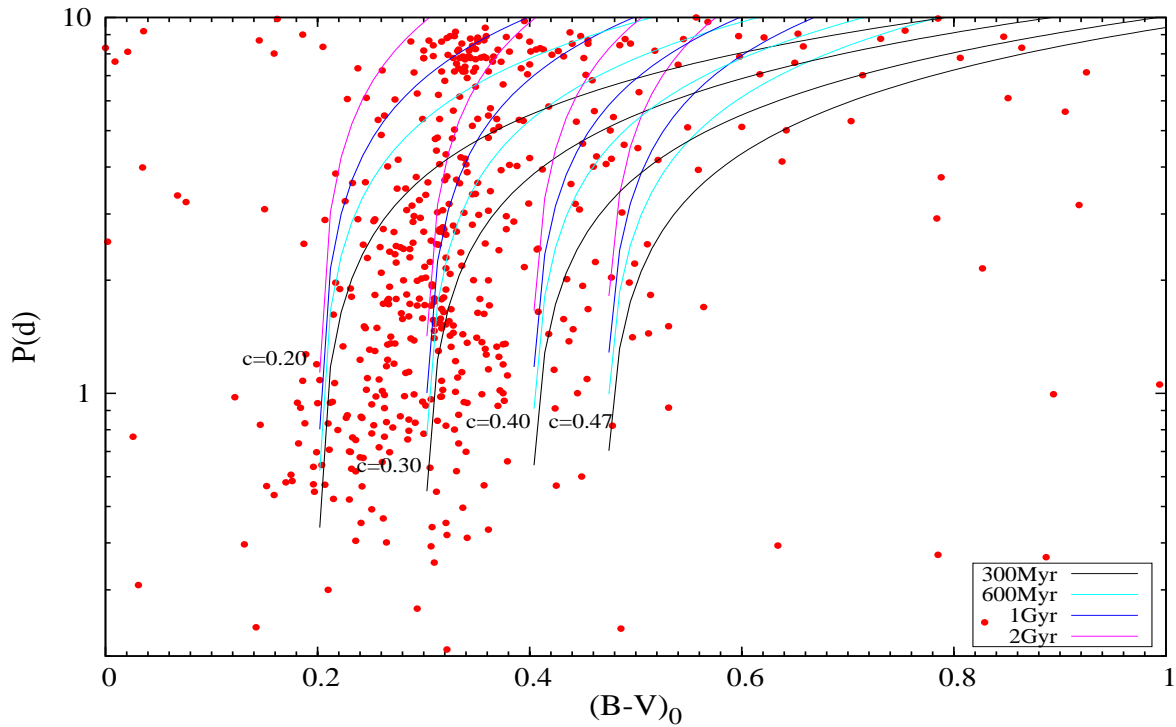}
\caption{Rotation period $P$ (in d) of our sample stars as a function of \bvo. See text for the description of the plot. }
\label{Period_vs_Colour}
\end{figure*}

In Fig.\,\ref{Period_vs_Colour} we show the photometric period as a function of \bvo\ for all the stars that we classified as spotted stars. To transform $(g-r)_0$ to \bvo\ we used the equation \citep{Jester2005}:
\begin{equation}
(B-V)_0 = 0.98(g-r)_0 + 0.22
\end{equation}
The figure clearly shows a correlation between rotational period and color index. There is a clear trend for hotter stars (especially for \bvo\ between 0.2 and 0.4) to have shorter rotation periods. This is in agreement with many past studies (see, e.g., \citet{Bouvier2013} for a review). According to \citet{Barnes2007}, the period $P$ is related to the colour \bvo\ and age $t$ by a relation of the form $P = t^n \times a[(B-V)_0 - c]^b$ where $a$, $b$, $c$, and $n$ are called the gyrochronology parameters. For $n$, a value of 0.5 is usually assumed, following the findings of \citet{Skumanich1972}. The most recent determinations for the other gyrochronology parameters are given by \citet{Meibom2009ApJ...695..679M}, who found $a = 0.770 \pm 0.014$, $b = 0.553 \pm 0.052$, and $c = 0.472 \pm 0.027$. We will take the expression
\begin{equation}
P = 0.77 \sqrt{t} ((B-V)_0 - 0.47) ^{0.55}
\end{equation}
as the starting point of our investigation, as this specific value of c = 0.47 has gained some status as a canonical value (cf. \citealt{Barnes2010}). The problem with this function is that it is not defined for \bvo\,$<$\,0.47 and therefore does not adequately describe the observations, as there is an unambiguous extension of a strong correlation of rotation period with colour index at least down to $(B-V)_0 =  0.17$. Our plot in Fig.\,\ref{Period_vs_Colour} is modelled on figure 6 in \citet{Reinhold2015} (RG15 in what follows). In that paper, the authors work with a value of c (see above) equal to 0.4, quoting \citet{Barnes2007}. The asymptote at c = 0.4 is also clearly indicated in figure 6 of RG15. \citet{Barnes2007} actually uses c = 0.4 as well as c = 0.5 in his paper, according to which, the sample of stars are being fitted. For reasons stated in section 2 of this paper, we have limited our study to stars with rotation periods below 10 days. That difference notwithstanding, our Fig.\,\ref{Period_vs_Colour} shows both an important agreement and an important difference with figure 6 in RG15: We see the same sparse population of stars with shorter periods at high values of \bvo\ as they do; however, the clear and sharp break to the left of the \bvo\,=\,0.47 line in RG15 is not seen at all in our data. This is an intriguing finding and needs further exploration.

In order to examine the nature of this observed extension of the canonical rotation period-colour relation, we adapted equation (7) to allow for thresholds in the colour lower than \bvo\,=\,0.47. We plotted curves of the form $P = 0.77 \sqrt{t} ((B-V)_0 - c) ^{0.55}$ with $c$ taking the values 0.2, 0.3 and 0.4 respectively, along with equation (7) itself (i.e. with c = 0.47), onto Fig.\,\ref{Period_vs_Colour}. These values were chosen arbitrarily, as a first exploration of the nature of a continuation of the the period-colour relation to hotter stars. For each chosen value of $c$, we show plots for ages of 300~Myr, 600~Myr, 1~Gyr and 2~Gyr.

We make a number of interesting observations based on Fig.\,\ref{Period_vs_Colour}:

i)  instead of the sharp cut-off at \bvo\,=\,0.47 seen in RG15, there is a similar cut-off in Fig.\,\ref{Period_vs_Colour}, at \bvo\,$\approx$\,0.17. This is not the first time this has been seen. In an independent study, \citet{Balona2013b} generated a similar feature in figure 7 of that paper. Although only a few dozen stars in that figure lie blueward of \bvo\,=\,0.47, the trend is unmistakable.

ii) note the sparse population of stars blueward of \bvo\,=\,0.17. Since we sampled stars from A0 to cooler temperatures, we expect the colour domain shown in Fig.\,\ref{Period_vs_Colour} to be populated down to \bvo\,=\,0.0. The density of stars in Fig.\,\ref{Period_vs_Colour} falls off quite suddenly below \bvo\,=\,0.17. We have no explanation for this. In RG15, the \bvo\ domain does not extend blueward of 0.15, so we can not make a comparison of this feature with their results.

iii)  Contrary to expectation, there is no preference for higher rotation velocities for the stars blueward of \bvo\,=\,0.17 in Fig.\,\ref{Period_vs_Colour}. Rather, the stars seem to possess rotation speeds evenly scattered between 0.3 d and 10 d (our upper limit). Admittedly, only few dozen stars appear blueward of \bvo\,=\,0.17 in our figure, too small a number to make a definitive statement.

We wish to emphasise the conservative approach we have taken in selecting candidate stars showing rotational modulation - as discussed earlier in this section. This gives us confidence in the reality of the break around \bvo\,=\,0.17. The most intriguing, and potentially most important feature of the distribution of the 513 stars plotted in Fig.\,\ref{Period_vs_Colour}, is that the relation between rotation period and intrinsic colour is neither linear nor random, but appears to follow the same Skumanich-type law that has been associated with the stars redward of \bvo\,=\,0.47 in the past (as discussed in detail in \citet{Reinhold2015}). This relation is clearly visible even in the absence of the exploratory curves plotted onto Fig.\,\ref{Period_vs_Colour}. The most reasonable conclusion we can draw from this, is that our results, in agreement with the results independently presented in \citet{Balona2013b}, indicate an extension of the canonical relation between rotation period and intrinsic colour, blueward of the canonical cut-off.

It is interesting to note that, even in RG15, as well as in figure 8 of RRB13, a clear relation between rotation period and colour is also seen for a population of stars blueward of \bvo\,=\,0.4 : In RRB13 this population extends to \bvo\,=\,0.3 and in RG15 it extends to \bvo\,=\,0.2. The large scatter of these populations for any fixed value of \bvo\ appears more akin to the classical Skumanich age-dependent scatter than to a simple decline of surface rotation speed with decreasing mass for stars blueward of F5 (the canonical view). We have now detected a similar mode of behaviour blueward of \bvo\,=\,0.4. Note that the stars plotted in Fig.\,\ref{Period_vs_Colour} are limited to the new rotational variables we discovered in the Kepler sample, which is why we do not reproduce the large populations of points redward of \bvo\ seen in RRB13 and RG15.



The very clear presence of a correlation between rotation period and colour index down to \bvo\,=\,0.17 is a much stronger confirmation of the trend already noted in \citet{Balona2013b, Balona2013a} for the clusters NGC\,6866 and NGC\,6819. It was assumed that the low-frequency variation seen in the {\it Kepler} observations of these stars is the rotational frequency, as found in cooler stars. This fact is supported by the agreement between the distribution of equatorial velocities resulting from the assumption and the distribution of equatorial velocities derived from line broadening \citep{Balona2011c}.
The trend displayed in \citet{Balona2013b, Balona2013a}, where the period-colour correlation extends to \bvo\,=\,0.17, is substantially
strengthened by the results we show in Fig.\,\ref{Period_vs_Colour}.
This clear correlation needs explanation. 
Since there is less brightness difference between spot and photosphere for hotter stars, leading to lower photometric amplitudes, it is very difficult to observe rotational modulation in stars earlier than mid-F type, from the ground. Hence, for normal A type stars with \bvo\,$<$\,0.3, rotational periods derived from rotational modulation are not available in the literature.

In fact, the empirical function used by \citet{Barnes2007} is undefined for \bvo\,$<$\,0.40 and the rotational evolution model developed by \citet{Barnes2010} is not applicable for these stars. Our study as well as some earlier studies suggests that there might be no definitive break in the relationship at the boundary between convective and radiative photospheres as previously suggested by \citet{Barnes2003}. We conclude that some additional mechanisms must also play a role in the PtM relationship. Hence, we reiterate the observation by \citet{Balona2013b}, that a re-evaluation of the physical cause of the PtM relationship (or a re-evaluation of the presence of convection in A-type stars) is required. 
\section{Conclusion} \label{sec:concl}

The time-series photometry from the {\it Kepler} archive at MAST was used to investigate solar-like oscillations, pulsations and rotational behaviour of A-K stars in the {\it Kepler} field. 
This study builds on prior work done on solar-like oscillations in red giants in open clusters and in the {\it Kepler} field in general \citep{Stello2009b, Stello2010, Basu2010, Huber2014}. We independently identified solar-like oscillations in 23 new red giants. The basic astrophysical parameters such as mass, radius and luminosity for these stars were determined using their pulsational properties. On comparison with the luminosities and radius from GAIA DR2, we found that the GAIA-based results agree well with our estimated values from the solar-like oscillations.

Observations from space are changing the outlook of $\gamma$\,Dor and $\delta$\,Sct variables. In the present study, a global analysis of a sample of 15\,106 stars has been performed to search for non-radial pulsators and hybrids. A careful seismic study of individual stars is needed to confirm their classification and fully characterize their properties and understand the relationship between them. In this regard, our list of candidate non-radial pulsators will serve as an useful resource, for any further detailed analysis of individual objects. 
The $\gamma$\,Dor stars are difficult to distinguish from rotational variability caused by starspots, but in some stars the periodogram morphology can help to distinguish between the two scenarios. By carefully looking for the presence or absence of harmonics, we attributed our targets respectively to the class of rotationally variable stars (those stars showing harmonics of the rotation period in their periodograms) or to the class of the $\gamma$\,Dor stars (those without harmonics). 

Until recently, the $\gamma$\,Dor stars were not considered very useful as asteroseismic tools, owing to the unknown effects of rotation on the frequencies and the lack of mode identification. A study by \citet{Balona2011a} suggested that pulsation and rotation periods might be very closely related to each other. Recently, \citet{Ouazzani2017} investigated the effect of uniform rotation on the g-mode pulsation spectra, and more specifically, on the g-modes period spacings. It led to the introduction of a new observable $\Sigma$. This is the slope of the period spacing when plotted as a function of period. It is based on the non-uniformity of the period spacings of $\gamma$Dor stars which is related to the internal rotation. They further  studied the relationship between this new observable and different stellar parameters, such as metallicity, centrifugal distortion and type of mixing. Many of the newly discovered $\gamma$\,Dor stars in our study can be suitable targets for a similar analysis, but also including differential rotation. Moreover, with missions such as TESS and PLATO, $\gamma$\,Dor variables will provide a useful testing bed for this new observable and asteroseismology in general, because it needs to be further investigated to what extent the differential rotation influences the excitation of the observed modes.

In this study, several $\delta$\,Sct pulsators were identified and further study on identification of modes by means of ground-based multi-colour photometry will be vital to reveal the potential for asteroseismology in these stars. To firmly characterize the pulsation properties of the candidate $\gamma$\,Dor/$\delta$\,Sct hybrid stars, a more detailed analysis, also involving a spectroscopic study, is required. We studied the two new empirical observables of $Energy$ and $Efficiency$, first introduced by \citet{Uytterhoeven2011}, for a larger sample of non-radial pulsators. The peak values of the distributions in $Energy$, $Efficiency$ and $Amplitude$ for $\delta$\,Sct, $\gamma$\,Dor/$\delta$\,Sct hybrid and $\gamma$\,Dor variables are presented according to the gaussian fits to the histograms. The peaks vary in the \logen\ distribution by more than two orders of magnitude, where the $\delta$\,Sct stars have the higher values. Due to this, we suggest that the distribution in the logarithm of $Energy$ (\logen) can be used as a potential tool to distinguish the non-radial pulsators, to some extent, along with some other classification scheme. Further investigations are required to understand the physical basis of these empirical observables and its impact on the relationship between $\delta$\,Sct, $\gamma$\,Dor and hybrid variables.

A potentially very significant result of the present work is the detection of a correlation of rotation period with colour index of {\it Kepler} field stars down to \bvo\,=\,0.2 or even lower. This correlation is based on the measurement of rotation periods for 513 stars which are likely to be rotational variables owing to the presence of harmonics of the dominant frequency in the periodogram (or peak periods too long to be attributed to $\gamma$\,Dor pulsations). There is no known explanation for this observation. We did detect the expected correlation between rotational period and colour index, which is explained as the result of magnetic braking of rotation via the interaction of convective envelopes with stellar rotation in cooler stars and which is generally applied to stars with \bvo\,$>$\,0.47 and used in the technique known as 'gyrochronology' \citep{Barnes2001,Barnes2003,Barnes2007, Barnes2010}. We were surprised, however, to find an additional correlation for hotter stars, mimicking the established relation to some extent.

It is believed that for stars hotter than mid-F, rotational braking is negligible [hence, the limit of applicability of the PtM relation to \bvo\,$>$\,0.47]. If the cause of the PtM relationship is rotational braking due to mass-loss arising from convection, one is led to conclude that mass-loss and convection are important in A and early F-type stars as well. However, the convective zones in the envelopes of these hot stars are believed to be very thin. Hence, we conclude that either the role of convection in A-type stars is not fully understood or that something else is responsible for, at least, a period-colour relationship very similar to the standard PtM relationship, but for hotter stars.

\section{Acknowledgments}

The work presented here is carried out under the international projects, namely, Indo-South African DST/INT/SA/P-02 \& INT/SAFR/P-3(3)/2009 and Indo-Belgium ‘BINA’ projects DST/INT/Belg/P-02 \& BL/11/IN07. SC wishes to thank the Aryabhatta Research Institute of Observational Sciences for their kind hospitality and acknowledges the stipend received from Indo-Russian RFBR project INT/RFBR/P-118, to carry out this work. SC acknowledges partial financial support from the Polish NCN grant 2015/18/A/ST9/00578. We are grateful to Prof. Luis A. Balona, Prof. Patricia J. Lampens, Prof. Gerald Handler, Dr. Blesson Mathew, and the anonymous referee for their suggestions and comments to improve this paper. This paper includes data collected by the {\it Kepler} mission. Funding for the {\it Kepler} mission is provided by the NASA Science Mission directorate. The authors wish to thank the {\it Kepler} team for their generosity in allowing the data to be released and for their outstanding efforts which have made these results possible. Much of the data presented in this paper were obtained from the Mikulski Archive for Space Telescopes (MAST). STScI is operated by the Association of Universities for Research in Astronomy, Inc., under NASA contract NAS526555. Support for MAST for non-HST data is provided by the NASA Office of Space Science via grant NNX09AF08G and by other grants and contracts.
\bibliography{sowgata_ms}

\end{document}